\documentclass{aa}
\usepackage{graphicx,psfrag,amsmath}
\begin{document}
\title{High precision effective temperatures for 181 F--K dwarfs
       from line-depth ratios
\thanks{Based on spectra collected with the ELODIE spectrograph at the
1.93-m telescope of the Observatoire de Haute Provence (France).}
}
\titlerunning{Precise temperatures for 181 F--K dwarfs}
\author{V.V. Kovtyukh \inst{1},\, C. Soubiran \inst{2},\, S.I.
 Belik \inst{1},\, and N.I. Gorlova \inst{3}}
\authorrunning{Kovtyukh et al.}
\offprints{V.V. Kovtyukh, \\
           e-mail: val@deneb.odessa.ua}
\institute{
Astronomical Observatory of Odessa
National University and Isaac Newton Institute of Chile,
Shevchenko Park, 65014, Odessa, Ukraine\\
email: val@deneb.odessa.ua
\and
Observatoire de Bordeaux, CNRS UMR 5804, BP 89, 33270 Floirac, France
\and
Steward Observatory, The University of Arizona,
Tucson, AZ, USA 85721
}
\date{Received 6 May 2003; accepted }
\abstract{Line depth ratios measured on  high resolution
($R$=42\,000), high $S/N$ echelle spectra are used for the
determination of precise effective temperatures of 181 F,G,K Main
Sequence stars with  about solar metallicity (--0.5 $<$ [Fe/H]
$<$ +0.5). A set of 105 relations is obtained which rely $T_{\rm
eff}$ on ratios of the strengths of lines with high and low
excitation potentials, calibrated against previously published
precise (one per cent) temperature estimates. The application
range of the calibrations is 4000--6150 $K$ (F8V--K7V). The
internal error of a single calibration is less than 100 $K$, while
the combination of all calibrations for a spectrum of $S/N$=100
reduces uncertainty to only 5--10 $K$, and for $S/N$=200 or higher
-- to better than 5 $K$. The zero point of the temperature scale
is directly defined from reflection spectra of the Sun with an
uncertainty about 1 $K$. The application of this method to
investigation of the planet host stars properties is discussed.
\keywords{Stars: fundamental parameters
 -- stars:temperatures -- stars:dwarfs -- planetary systems}
 }

\maketitle

\section{Introduction}
The determination of accurate effective temperatures is a
necessary prerequisite for detailed abundance analysis. In this
paper we focus on dwarfs with solar metallicity (--0.5 $<$ [Fe/H]
$<$ +0.5) to contribute to the very active research field
concerning the fundamental parameters of stars with planets. High
precision temperatures of such stars might help to resolve two
outstanding questions in the extra-solar planetary search. Namely,
to get a definite confirmation of the metal richness of the stars
that harbor planets, and secondly, perhaps to rule out some
low-mass planetary candidates by detecting subtle variations in
the host's temperature due to star-spots. The numerous studies of
the large fraction of the known extra-solar planet hosts ($\sim$80
out of $\sim$100 known systems) have revealed their larger than
average metal richness (Gonzalez 1997; Fuhrmann, Pfeiffer \&
Bernkopf 1998; Gonzalez et al. 2001 and references therein; Takeda
et al 2001; Santos et al. 2003 and references therein). The
reliability of this result depends mainly on the accuracy of the
model atmosphere parameters, with effective temperature ($T_{\rm
eff}$) being the most important one.

The direct method to determine the effective temperature of a star
relies on the measurement of its angular diameter and bolometric
flux. In practice certain limitations restrict the use of this
fundamental method to very few dwarfs. Other methods of
temperature determination have errors of the order 50--150 $K$,
which translates into the [Fe/H] error of $\sim$0.1 dex or larger.
The only technique capable so far of increasing this precision by
one order of magnitude, is the one employing ratios of lines with
different excitation potentials $\chi$. As is well known, the
lines of low and high $\chi$ respond differently to the change in
$T_{\rm eff}$. Therefore, the ratio of their depths
$r=R_{\lambda1}/R_{\lambda2}$ (or equivalent widths, EW) should be
very sensitive temperature indicator. The big advantage of using
line-depth ratios is the independence on the interstellar
reddening, spectral resolution, rotational and microturbulence
broadening.

The reader is referred to Gray (1989, 1994) and Gray \& Johanson
(1991) to learn more about the history and justification of the
line ratio method. Applying this method to the Main-Sequence (MS)
stars, they achieved precision as high as 10 $K$. The most recent
works on the subject are by Caccin, Penza \& Gomez (2002) who
discuss the possible weak points of this technique for the case of
dwarfs (see below), and the fundamental contribution by
Strassmeier \& Schordan (2000) who report 12 temperature
calibrations for giants with an error of only 33 $K$.

So far however the line-ratio method has not been fully utilized
for the purposes other than just temperature estimation by itself.
One of few applications is the chemical abundance analysis of
supergiants, where it has proved the anticipated high efficiency
and accuracy. Thus, Kovtyukh \& Gorlova (2000, hereafter Paper I)
using high-dispersion spectra, established 37 calibrations for the
temperature determination in supergiants (a further study
increased this number to 55 calibrations). Based on this
technique, in the series of 3 papers Andrievsky et al.
(\cite{andet02} and references therein) derived temperatures for
116 Cepheids (from about 260 spectra) at a wide range of
galactocentric distances (R$_{g}$=5--15 kpc) with a typical error
5--20 $K$. The high precision of this new method of temperature
determination allowed them to uncover the fine structure in the
Galactic abundance gradients for many elements. Even for the most
distant and faint objects ($V \simeq$  13--14 mag) the mean error
in $T_{\rm eff}$ was no larger than 50--100 $K$, with maximum of
200 $K$ for spectra with lowest $S/N(=$40--50).

Another example concerns T Tau stars. For young stars,
uncertainties in reddening due to variable circumstellar
extinction invalidate the photometric color method of effective
temperature determination. Using 5 ratios of FeI and VI lines
calibrated against 13 spectral standards, Padgett
(\cite{padgett96}) determined the effective temperature of 30 T
Tau stars with a 1 $\sigma$ uncertainty lower than 200 $K$.

The intent of this paper is to improve this technique, based on
our experience of applying it to supergiants (Paper I and
following publications), and expand it to the MS stars.  The wide
spectral range of ELODIE echelle spectra allowed to select many
unblended lines of low and high excitation potentials thus
improving the internal consistency of the method, whereas the
large intersection between the ELODIE database and published
catalogues of effective temperatures allowed to take care of
systematic effects. We obtained a median precision of 6 $K$ on
$T_{\rm eff}$ derived for an individual star. The zero-point of
the scale was directly adjusted to the Sun, based on 11 solar
reflection spectra taken with ELODIE, leading to the uncertainty
in the zero-point of about 1 $K$.

Temperature determined by the line ratio method may now be considered as
one of the few fundamental
stellar parameters that have been measured with
internal precision of better than 0.2\%.

\section{Observations and temperature calibrations}

The investigated spectra are part of the library collected with
the ELODIE spectrometer on the 1.93-m telescope at the
Haute-Provence Observatory (Soubiran et al. \cite{soubet98},
Prugniel \& Soubiran \cite{prusou01}). The spectral range is
4400--6800 \AA\AA\, and the resolution is $R$=42000. The initial
data reduction is described in Katz et al. (\cite{katzet98}). All
the spectra are parametrized in terms of $T_{\rm eff}$, logg,
[Fe/H], either collected from the literature or estimated with the
automated procedure TGMET (Katz et al. \cite{katzet98}). This
allowed us to select a sample of spectra of FGK dwarfs in the
metallicity range --0.5$ <$ [Fe/H] $<$ +0.5. Accurate Hipparcos
parallaxes are available for all of the stars of interest enabling
to determine their absolute magnitudes M$_{V}$ that range between
2.945 (HD81809, G2V) and 8.228 (HD201092, K7V). All the selected
spectra have a signal to noise ratio greater than 100. Further
processing of spectra (continuum placement, measuring equivalent
widths, etc.) was carried out by us using the DECH20 software
(Galazutdinov \cite{gal92}). Equivalent widths EWs and depths
$R_{\lambda}$  of lines were measured manually by means of a
Gaussian fitting. The Gaussian height was then a measure of the
line depth. This method produces line depths values that agree
nicely with the parabola technique adopted in Gray (1994). We
refer the reader to Gray (1994, and references therein),
Strassmeier \& Schordan (2000) for a detailed analysis of error statistics.

\begin{figure*}[hbtp] 
\begin{center}
\includegraphics[width=16cm]{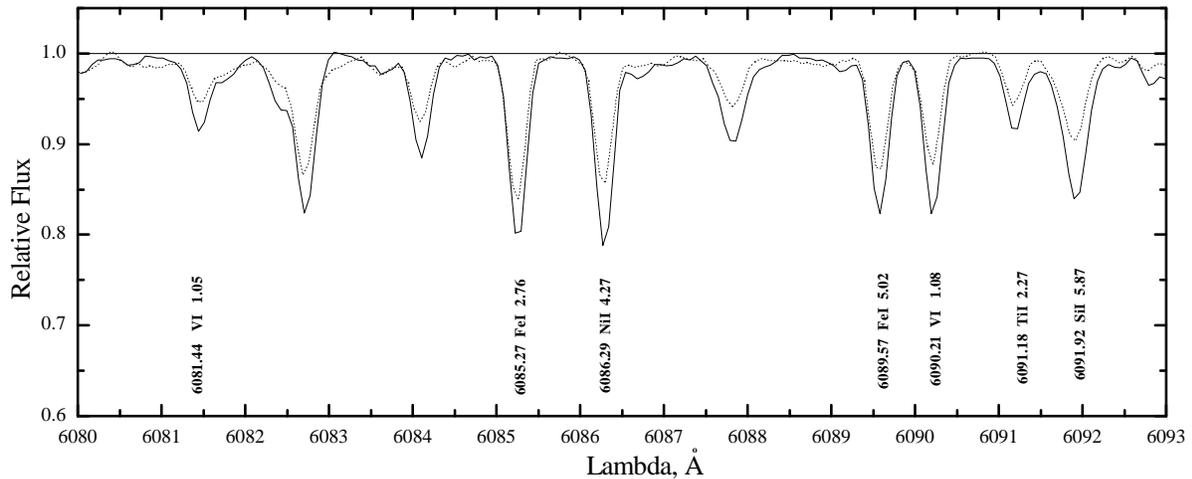}
\caption{Comparison spectra for two stars: $solid$ $line$ -- a
          planet-host star HD 217014
          (51 Peg), and $dotted$ $line$ -- a non-planet star HD 5294.
          Within the limits of the errors, both stars have identical
          temperatures (5778 and 5779 $K$, respectively), but different
          metallicities. Spectral lines used in temperature
          calibrations are identified at the bottom with their
          wavelength, element, and lower excitation potentials $\chi$ in eV.
          We used ratios 6081.44/6089.57, 6085.27/6086.29, 6085.27/6155.14,
          6089.57/6126.22, 6090.21/6091.92, 6090.21/6102.18, 6091.92/6111.65
          and others.}
\end{center} 
\end{figure*}

 Following Caccin's, Penza \& Gomez (2002) results, where a careful analysis
of the anticipated problems for the Solar-type stars has been carried out,
we did not use ion lines and high-ionization elements (like C, N, O)
due to their strong sensitivity to gravity.

  Gray (1994) showed that the ratio of lines VI 6251.82 and FeI
6252.55 depends strongly on metallicity. The reason is that the
strong lines like FeI 6252.55 (R$_{\lambda}$=0.52 for the Sun) are
already in the dumping regime, where the linearity of EW on
abundance breaks down. In addition, as was shown in the careful
numerical simulations by Stift \& Strassmeier (1995), this ratio
(of 6251.82 and 6252.55 lines) is also sensitive to rotational
broadening. Significant effects were found for v$sin$i as small as
0--6 km s$^{-1}$ (for solar-like stars). We therefore avoided to use
strong lines in our calibrations. Indeed, Gray (1994) concluded
that, as expected, the weak-line ratios are free from the effects
of metallicity. As to the effect of rotation, we should note that
all objects in our sample are old Main Sequence stars with slow to
negligible rotation (v$sin$i$<$15 km s$^{-1}$), which is
comparable to the instrumental broadening.

  Thus, we initially selected about 600 pairs of 256 unblended
SiI, TiI, VI, CrI, FeI, NiI lines with high and low excitation
potentials within the wavelength interval 5300--6800 \AA\AA.

 These lines have been selected according to the following
criteria:

 (1) the excitation potentials of the lines in pair must differ as
     much as possible;

 (2) the lines must be close in wavelength; it turned out though
 that calibrations based on widely spaced lines (including from different
 orders) show same small dispersion as the closely spaced lines.
 Therefore, we retained all pairs with difference in wavelength up
 to 70 \AA ($\lambda_{2} - \lambda_{1}<$70 \AA);

 (3) the lines must be weak enough to eliminate a possible dependence
     on microturbulence, rotation and metallicity;

 (4) the lines must be situated in the spectral regions
     free from telluric absorption.

The next step was to choose the initial temperatures for
interpolation. This is a very important procedure since it affects
the accuracy of the final temperature scale,  namely, the run
of the systematic error with $T_{\rm eff}$ (Fig. 3). There is an
extended literature on MS stars temperatures. For 45 stars from
our sample (see Table 1) we based the initial temperature
estimates on the following 3 papers: Alonso, Arribas \&
Martinez-Roger (\cite{alonso96}, hereafter AAMR96), Blackwell \&
Lynas--Gray (\cite{black98}, hereafter BLG98) and DiBenedetto
(\cite{dib98}, hereafter DB98). In these works the temperatures
have been determined for a large fraction of stars from our sample
with a precision of about 1\%. AAMR96 used the Infrared Flux
Method (IRFM) to determine $T_{\rm eff}$ for 475 dwarfs and
subdwarfs with a mean accuracy of about 1.5\% (i.e., 75--90 $K$).
BLG98 also have determined temperatures for 420 stars with
spectral types between A0 and K3 by using IRFM and achieved an
accuracy of 0.9\%. DB98 derived $T_{\rm eff}$ for 537 dwarfs and
giants by the empirical method of surface brightness and Johnson
broadband $(V-K)$ color, the accuracy claimed is $\pm1$\%.
Whenever 2 or 3 estimates were available for a given star, we
averaged them with equal weights. These temperatures served as the
initial approximations for our calibrations.

First, for the above mentioned 45 stars with previously accurately
determined $T_{\rm eff}$ we plotted each line ratio against
$T_{\rm eff}$, and retained only those pairs of lines that showed
unambiguous and tight correlation. We experimented with a total of
nearly 600 line ratios but adopted only the 105 best - the ones
showing the least scatter. These 105 calibrations consist of
92 lines, 45 with low ( $\chi<$2.77 eV ) and 47 with high (
$\chi>$4.08 eV ) excitation potentials. Judging by the small
scatter in our final calibrations (Fig.2) and $T_{\rm eff}$, the selected
combinations are only weakly sensitive to effects like rotation,
metallicity and microturbulence. This confidence is reinforced by
the fact that the employed lines belong to the wide range of
chemical elements, intensity and atomic parameters, therefore one
can expect the mutual cancellation of the opposite effects.

Each relationship was then fitted with a simple analytical
function. Often calibrations show breaks which can not be
adequately described even by a 5th-order polynomial function (see
Fig. 2). Therefore, we employed other functions as well, like
Hoerl function ($T_{\rm eff}$=$ab^{r}*r^{c}$, where
$r=R_{\lambda1}/R_{\lambda2}$, $a,b,c$ -- constants), modified
Hoerl ($T_{\rm eff}$=$ab^{1/r} r^{c}$), power low ($T_{\rm
eff}$=$ar^{b}$), exponential ($T_{\rm eff}$=$ab^{r}$) and
logarithmic ($T_{\rm eff}$=a+b $ln(r)$) functions. For each
calibration we selected function that produced the least square
deviation. As a result, we managed to accurately approximate the
observed relationships with a small set of analytic expressions.
This first step allowed to select 105 combinations, with an rms of
the fit lower than 130 $K$, the median rms being 93 $K$. Using
these initial rough calibrations, for each of the 181 target stars
we derived a set of temperatures (70--100 values, depending on the
number of line ratios used), averaged them with equal weights, and
plotted these mean $T_{\rm eff}$ (with errors of only 10--20 $K$)
versus line ratios again, thus determining the  preliminary
calibrations (for which the zero-point had yet to be adjusted).

We would like to point out that the precision of our calibrations
varies with temperature. In particular,
at high $T_{\rm eff}$ the lines with low $\chi$ become very weak causing
line depth measurement to be highly uncertain.
Therefore, for each calibration we determined the
optimum temperature range where the maximum accuracy is attained
(no worse than 100 $K$), so that for a given star only
a subset of calibrations can be applied.

\begin{figure}[hbtp] 
\begin{center}
\includegraphics[width=9cm]{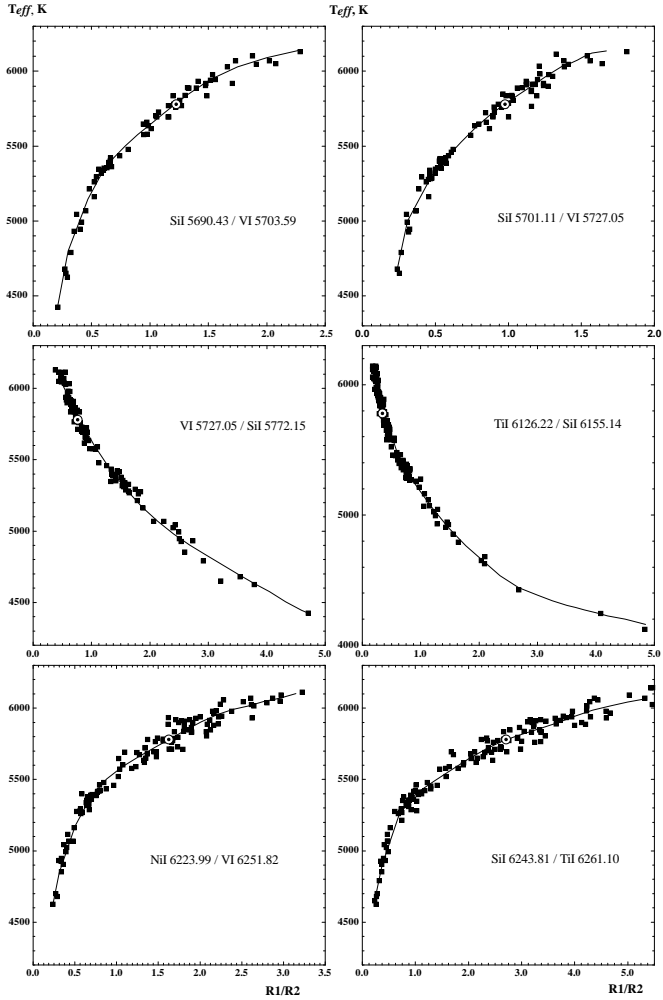} 
\caption{Our
         final calibrations of temperature versus line depth ratios
         $r$=R1/R2. The temperatures are shown as the average value derived
         from all calibrations
         available for a given star. The errors in temperature are less
         than the symbol size. The typical error in line ratio is 0.02--0.05.
         Position of the Sun is marked by the standard symbol.}
\end{center} 
\end{figure}

\begin{figure}[hbtp] 
\begin{center}
\includegraphics[width=6cm]{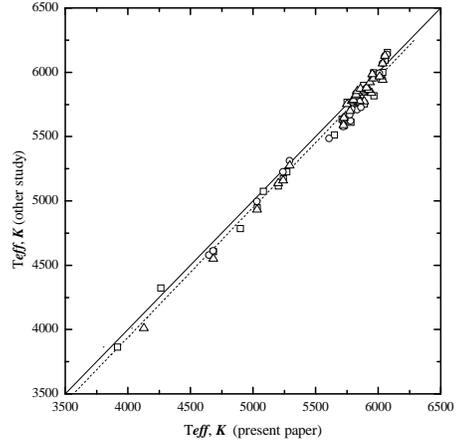}
\caption{Comparison between the temperatures derived in the present work and
 those derived by  AAMR96 -- $squares$,   BLG98 -- $circles$, and DB98 --
 $triangles$.
 The dashed line represents the linear fit to the data, and the solid line
 represents the one-to-one correlation}
\end{center} 
\end{figure}

\begin{figure}[hbtp] 
\begin{center}
\includegraphics[width=8cm]{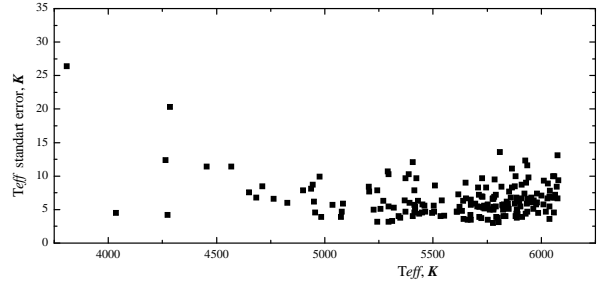}
\caption{Standard error of the mean versus effective
          temperature averaged over all available line ratios.}
\end{center} 
\end{figure}

What are the main sources of random errors in the line ratio
method? The measurement errors in line depths are mainly caused by
the continuum placement uncertainty and by the Gaussian
approximation of the line profile. In addition, the individual
properties of the stars, such as metallicity, spots, rotation,
convection, non-LTE effects, and binarity may also be responsible
for the scatter observable in Fig. 2. The detailed analysis of
these and other effects can be found in Paper I, Strassmeier \&
Schordan \cite{ss00} and in works by D.F.Gray. We estimate that
the typical error in the line depth measurement
$r=R_{\lambda1}/R_{\lambda2}$ is 0.02-0.05, implying an error in
temperature of about 20--50 $K$.

The mean random error of a single calibration is 60--70 $K$
(40--45 $K$ in the most and 90--95 $K$ in the least accurate
cases).

The use of $\sim$70--100 calibrations reduces the uncertainty to
5--7 $K$ (for spectra with $S/N$=100--150). Better quality spectra
($R>$100,000, $S/N>400$) should in principle allow the uncertainty
of just 1--2 $K$. Clearly, time variation of the temperature for a
given star should be readily detected by this method, since the
main parameters that cause scatter due to star-to-star
dissimilarities ( gravity, rotation, [Fe/H], convection,
non-LTE effects etc.) are fixed for a given star. The temperature
variation of several degrees in mildly active stars may be
produced by the surface features and rotational modulation, as for
example has been documented for the G8 dwarf $\xi$ Bootis A (Toner
\& Gray \cite{tongr88}) and $\sigma$ Dra (K0V, Gray et al.
\cite{gray92}).

The next stage is to define the zero point of our temperature
scale. Fortunately, for dwarfs (unlike for supergiants) a
well-calibrated standard exists, the Sun. Using our
preliminary calibrations and 11 independent solar spectra from
the ELODIE library (reflection spectra of the Moon and asteroids),
we obtained a mean value of 5733 $\pm 0.9 K$ for the Sun's
temperature. Considering Sun a normal star, we adjusted our
calibrations by adding 44 $K$ to account for the offset between
the canonical Solar temperature of 5777 $K$ and our estimate. The
possible reasons for this small discrepancy are discussed below.

\begin{table}
\begin{center}
\caption[]{RMS of a linear regression between Teff and Str\"
omgren $b-y$ using effective temperatures obtained by other
authors and in this study with N common stars.}
\begin{tabular}{lccc}
\hline
\hline
author  &N & $\sigma_{\rm others}$ K &  $\sigma_{\rm our}$ K \\
\hline
AAMR96  &  30 & 102 & 78 \\
BLG98   &  25 &  71 & 65 \\
DB98    &  29 & 113 & 87 \\
EDV93   &  30 &  63 & 29 \\
\hline
\end{tabular}
\end{center}
\end{table}

Another point concerns the difference between the zero-point of
our temperature scale and that of other authors. Comparing 30
common objects, we find that AAMR96 scale underestimates
temperatures by 45 K near the solar value compared to ours, but
apart from that, the deviations are random, no trend with $T_{\rm
eff}$ is present. The 45 $K$ offset may arise from the various
complications associated with observing the Sun as a star, and/or
problems in the models used by AAMR96, like underestimation of
convection in the grid of the model atmosphere flux developed by
Kurucz. After correcting AAMR96 zero-point for the 45 K offset,
the mean random error of their scale becomes 65 $K$ (where we
neglect the error of our own scale which is an order less).

The temperatures of BLG98 are also in a good agreement with our
estimates -- except for a 48 $K$ offset, no correlation of the
difference with temperature is observed. The mean dispersion is 63
$K$ (for 26 common stars), which is within the errors of BLG98
scale.

Comparing with DB98: for the 29 star in common, their temperatures
are on average 41 $K$ below ours, and the mean error is $\pm53 K$.

Thus, the temperatures derived in AAMR96, BLG98 and DB98 have good
precision, though the absolute values are somewhat low relative to
the Sun. The reason may be due to the difficulty of the
photometric measurements of the Sun, as well as indicate some
problems in the model atmosphere calculations employed. For
example, the Sun's temperatures derived in AAMR96 and DB98 are
identical -- 5763 $K$, which is below the nominal value of 5777
$K$. Besides, the mean temperatures of solar analogue stars
(spectral types G2-G3, [Fe/H]$\approx$0.0, and Sun being of G2.5
type) derived in these papers, are significantly below the solar
value: 5720$\pm54 K$ (AAMR96, 3 stars), 5692$\pm31 K$ (BLG98, 11
stars) and 5702$\pm46 K$ (DB98, 7 stars). Our determination for
the G2-G3 spectral types is 5787$\pm14 K$, based on 12 stars. This
demonstrates that a small error (0.8\%) affects the zero point
of the IRFM method, because when applied to the Sun and the solar
type stars, it returns inconsistent results.

We also compared our estimates of $T_{\rm eff}$ with photometrical
temperatures.
EDV93 derived temperatures of 189 nearby field F,G disk
dwarfs using theoretical calibration of temperature versus Str\"{o}mgren
(b--y) photometry (see Table 1). The mean difference between $T_{\rm eff}$
of Edvardsson  et al.(\cite{edvet93}) and ours is only $-14 K$
($\sigma$=$\pm$67 $K$, based on 30 common stars).

 To compare our temperatures to Gray (1994), we used his
calibration of (B-V)$_{corr}$ corrected for metallicity. Our scale is +11 K
lower ($\sigma$=$\pm$61 $K$, 24 stars).

Summarizing, we demonstrated that our temperature scale is in
excellent agreement with the widely used photometric scales, while
both the IRFM method and the method of surface brightness predict
too low values for the temperature of the Sun and the solar type
stars.

Fig. 4 shows the sensitivity of our technique to temperature. Two
outliers with errors greater than 20 $K$ are the cold dwarfs
HD28343 and HD201092, known as flaring stars. For other stars the
internal errors range between 3 and 13 $K$, with a median of 6
$K$.

\begin{table*}
\caption[]{Program stars. Asterisks indicate stars with planets.}
\begin{tabular}{rrrcrrccccclc}
\hline
\hline
HD/BD &HR&Name&$T_{\rm eff}$&N&$\sigma$, K&$T_{\rm eff}$&$T_{\rm eff}$&
$T_{\rm eff}$&$T_{\rm eff}$& Mv & B--V& rem \\
 &  &  & this paper  & & &EDV93&AAMR96& BLG98 & DB98 & & & \\
\hline
  1562  & -- &            &5828& 97& 5.8&     &     &    &    &  5.006&0.585& \\
  1835  &  88&  9 Cet     &5790& 68& 5.5&     &     &5713&5774&  4.842&0.621& \\
  3765  & -- &            &5079& 87& 4.7&     &     &    &    &  6.161&0.954& \\
  4307  & 203& 18 Cet     &5889& 91& 5.0& 5809& 5753&    &5771&  3.637&0.568& \\
  4614  & 219& 24 Eta Cas &5965& 69& 6.4& 5946& 5817&    &    &  4.588&0.530& \\
  5294  & -- &            &5779& 86& 6.6&     &     &    &    &  5.065&0.610& \\
  6715  & -- &            &5652& 97& 6.7&     &     &    &    &  5.079&0.658& \\
  8574  & -- &         &6028& 61& 6.7&     &     &    &    &  3.981&0.535& * \\
  8648  & -- &            &5790& 59& 7.2&     &     &    &    &  4.421&0.643& \\
  9826  & 458& 50 Ups And &6074& 44&13.1& 6212& 6155&6136&  & 3.452&0.496& * \\
 10145  & -- &            &5673& 96& 4.2&     &     &    &    &  4.871&0.667& \\
 10307  & 483&            &5881& 94& 4.0& 5898& 5874&    &    &  4.457&0.575& \\
 10476  & 493& 107 Psc    &5242& 69& 3.2&  & 5172&5223&5157&  5.884&0.819& \\
 10780  & 511&            &5407& 95& 4.0&     &     &    &    &  5.634&0.767& \\
 11007  & 523&            &5980& 84& 7.4&     &     &    &    &  3.612&0.524& \\
 13403  & -- &            &5724& 91& 7.0&     & 5585&5577&5588&  3.949&0.616& \\
 13507  & -- &        &5714& 91& 5.4&     &     &    &    &  5.123&0.637& * \\
 13825  & -- &            &5705& 96& 5.5&     &     &    &    &  4.700&0.674& \\
 14374  & -- &            &5449& 77& 4.6&     &     &    &    &  5.492&0.757& \\
 15335  & 720& 13 Tri     &5937& 84& 6.6& 5857&     &    &5921&  3.468&0.539& \\
 17674  & -- &            &5909& 58& 8.7&     &     &5875&5880&  4.194&0.563& \\
 17925  & 857&         &5225& 87& 5.0&     &     &    &    &  5.972&0.864& \\
 18803  & -- &            &5659& 95& 3.5&     &     &    &    &  4.998&0.669& \\
 19019  & -- &            &6063& 56& 7.2&     &     &    &    &  4.445&0.508& \\
 19308  & -- &            &5844& 95& 5.4&     &     &    &    &  4.220&0.626& \\
 19373  & 937& Iot Per    &5963& 75& 5.1&     & 5996&5981&5951&  3.935&0.554& \\
 19994  & 962& 94 Cet     &6055& 56&10.0& 6104&   &    &    &  3.313&0.523& * \\
 22049  &1084& 18 Eps Eri &5084& 84& 5.9&  & 5076&    &    &  6.183&0.877& * \\
 22484  &1101&  10 Tau &6037& 60& 3.6& 5981& 5998&5944&5940&  3.610&0.527& \\
 23050  & -- &            &5929& 80& 9.0&     &     &    &    &  4.330&0.544& \\
 24053  & -- &            &5723& 93& 3.7&     &     &    &    &  5.183&0.674& \\
 24206  & -- &            &5633& 94& 4.8&     &     &    &    &  5.418&0.681& \\
 26923  &1322& V774 Tau   &5933& 77& 5.9&     &     &    &    &  4.685&0.537& \\
 28005  & -- &            &5980& 87& 6.1&     &     &    &    &  4.359&0.652& \\
 28099  & -- &            &5778& 85& 5.2&     &     &    &    &  4.747&0.660& \\
 28343  & -- &            &4284& 20&20.3&     &     &    &    &  8.055&1.363& \\
 28447  & -- &            &5639& 93& 6.3&     &     &    &    &  3.529&0.678& \\
 29150  & -- &            &5733& 89& 5.4&     &     &    &    &  4.934&0.668& \\
 29310  & -- &            &5852& 89& 7.7&     &     &5781&5775&  4.407&0.564& \\
 29645  &1489&            &6009& 57& 5.8& 6028&     &    &    &  3.504&0.548& \\
 29697  & -- &            &4454& 40&11.4&     &     &    &    &  7.483&1.108& \\
 30495  &1532& 58 Eri     &5820& 91& 5.7&     &     &    &    &  4.874&0.588& \\
\hline
\end{tabular}
\end{table*}

\begin{table*}
{Table 1 (Continued)}\\
\begin{tabular}{rrrcrrccccclc}
\hline
\hline
HD&HR&Name&$T_{\rm eff}$&N&$\sigma$, K&$T_{\rm eff}$&$T_{\rm eff}$&
$T_{\rm eff}$&$T_{\rm eff}$& Mv & B--V& rem \\
 &  &  & this paper  & & &EDV93&AAMR96& BLG98 & DB98 & & & \\
\hline
 30562  &1536&            &5859& 87& 6.8& 5886& 5822&5843&5871&  3.656&0.593& \\
 32147  &1614&            &4945& 65& 8.7&     &     &    &    &  6.506&1.077& \\
 34411  &1729& 15 Lam Aur &5890& 88& 4.3& 5889& 5847&5848&5859&  4.190&0.575& \\
 38858  &2007&            &5776& 81& 6.7&     &     &5669&5697&  5.014&0.584& \\
 39587  &2047&54 Chi1 Ori &5955& 71& 6.1& 5953&   &    &    &  4.716&0.545& \\
 40616  & -- &            &5881& 89&10.0&     &     &    &    &  3.833&0.585& \\
 41330  &2141&            &5904& 77& 5.5& 5917&     &    &    &  4.021&0.547& \\
 41593  & -- &            &5312& 92& 3.3&     &     &    &    &  5.814&0.802& \\
 42618  & -- &            &5775& 96& 6.6&     &     &    &    &  5.053&0.603& \\
 42807  &2208&            &5737& 81& 5.2&     &     &    &    &  5.144&0.631& \\
 43587  &2251&            &5927& 81& 4.4&     &     &    &    &  4.280&0.558& \\
 43947  & -- &            &6001& 82& 7.1& 5945&     &    &    &  4.426&0.507& \\
 45067  &2313&            &6058& 61& 4.6&     &     &    &    &  3.278&0.507& \\
 47309  & -- &            &5791& 95& 3.9&     &     &    &    &  4.469&0.623& \\
 50281  & -- &            &4712& 56& 8.5&     &     &    &    &  6.893&1.074& \\
 50554  & -- &            &5977& 77& 5.8&     &     &    &    &  4.397&0.529&*\\
 51419  & -- &            &5746& 94& 8.3&     &     &    &    &  5.013&0.600& \\
 55575  &2721&            &5949& 65& 6.6& 5963&     &    &5839&  4.418&0.531& \\
 58595  & -- &            &5707& 87& 8.3&     &     &    &    &  5.105&0.665& \\
 60408  & -- &            &5463& 97& 4.7&     &     &    &    &  3.100&0.760& \\
 61606  & -- &            &4956& 83& 4.6&     &     &    &    &  6.434&0.955& \\
 62613  &2997&            &5541& 90& 6.4&     &     &    &    &  5.398&0.695& \\
 64815  & -- &            &5864& 88& 8.3&     &     &    &    &  3.375&0.605& \\
 65874  & -- &            &5936& 85& 4.7&     &     &    &    &  3.100&0.574& \\
 68017  & -- &            &5651&100& 9.0&     & 5512&    &    &  5.108&0.630& \\
 68638  & -- &            &5430& 90& 6.3&     &     &    &    &  5.021&0.746& \\
 70923  & -- &            &5986& 82& 4.5&     &     &    &    &  3.879&0.556& \\
 71148  &3309&            &5850& 88& 5.1&     &     &    &    &  4.637&0.587& \\
 72760  & -- &            &5349& 91& 3.8&     &     &    &    &  5.628&0.796& \\
 72905  &3391& 3 Pi1 UMa  &5884& 79& 6.8&     &     &    &    &  4.869&0.573& \\
 73344  & -- &            &6060& 37& 6.8&     &     &    &    &  4.169&0.515& \\
 75318  & -- &            &5450& 78& 5.8&     &     &    &    &  5.345&0.717& \\
\hline
\end{tabular}
\end{table*}

\begin{table*}
{Table 1 (Continued)}\\
\begin{tabular}{rrrcrrccccclc}
\hline
\hline
HD&HR&Name&$T_{\rm eff}$&N&$\sigma$, K&$T_{\rm eff}$&$T_{\rm eff}$&
$T_{\rm eff}$&$T_{\rm eff}$& Mv & B--V& rem \\
 &  &  & this paper  & & &EDV93&AAMR96& BLG98 & DB98 & & & \\
\hline
 75732  &3522&55 Rho1 Cnc &5373& 97& 9.7&   &     &    &    &  5.456&0.851& * \\
 76151  &3538&            &5776& 88& 3.0& 5763&     &    &    &  4.838&0.632& \\
 76780  & -- &            &5761& 87& 5.0&     &     &    &    &  5.011&0.648& \\
 81809  &3750&            &5782& 85& 6.9&     & 5611&5619&    &  2.945&0.606& \\
 82106  & -- &            &4827& 76& 6.0&     &     &    &    &  6.709&1.000& \\
 86728  &3951& 20 LMi     &5735& 91& 5.6& 5746&     &    &    &  4.518&0.633& \\
 88072  & -- &            &5778& 82& 5.0&     &     &    &    &  4.717&0.593& \\
 89251  & -- &            &5886& 89& 6.3&     &     &    &    &  3.292&0.569& \\
 89269  & -- &            &5674& 95& 5.7&     &     &    &    &  5.089&0.645& \\
 89389  &4051&            &6031& 48& 8.9&     &     &    &    &  4.034&0.532& \\
 91347  & -- &            &5923& 75& 7.4&     &     &    &    &  4.725&0.513& \\
 95128  &4277& 47 UMa   &5887& 89& 3.8& 5882&     &    &    &  4.299&0.576& * \\
 96094  & -- &            &5936& 73&11.6&     &     &    &    &  3.725&0.550& \\
 98630  & -- &            &6060& 52&10.0&     &     &    &    &  3.043&0.553& \\
 99491  &4414& 83 Leo     &5509& 96& 8.6&     &     &    &    &  5.230&0.785& \\
101206  & -- &            &4649& 60& 7.6&     &     &4576&    &  6.750&0.983& \\
102870  &4540& 5 Bet Vir  &6055& 48& 6.8& 6176& 6095&6124&6127&  3.407&0.516& \\
107705  &4708& 17 Vir     &6040& 56& 7.8&     &     &    &    &  4.104&0.498& \\
108954  &4767&            &6037& 60& 5.5& 6060&     &6068&6068&  4.507&0.518& \\
109358  &4785& 8 Bet CVn  &5897& 72& 6.2& 5879& 5867&    &    &  4.637&0.549& \\
110833  & -- &            &5075& 80& 3.9&     &     &    &    &  6.130&0.938& \\
110897  &4845& 10 CVn     &5925& 68&12.3& 5795&     &    &5862&  4.765&0.510& \\
112758  & -- &            &5203& 83& 8.4&     & 5116&    &5137&  5.931&0.791& \\
114710  &4983& 43 Bet Com &5954& 71& 6.8& 6029& 5964&5959&5985&  4.438&0.546& \\
115383  &5011& 59 Vir     &6012& 40& 9.3& 6021&     &5989&5967&  3.921&0.548& \\
116443  & -- &            &4976& 83& 9.9&     &     &    &    &  6.175&0.850& \\
117043  &5070&            &5610& 98& 4.7&     &     &    &    &  4.851&0.729& \\
117176  &5072& 70 Vir  &5611&104& 4.7&     &     &5482&    &  3.683&0.678& * \\
119802  & -- &            &4763& 71& 6.6&     &     &    &    &  6.881&1.099& \\
122064  &5256&            &4937& 84& 8.1&     &     &    &    &  6.479&1.038& \\
122120  & -- &            &4568& 35&11.4&     &     &    &    &  7.148&1.176& \\
124292  & -- &            &5535& 89& 4.0&     &     &    &    &  5.311&0.721& \\
125184  &5353&            &5695& 89& 5.9& 5562&     &    &    &  3.898&0.699& \\
126053  &5384&            &5728& 79& 6.9&     &     &5635&5645&  5.032&0.600& \\
130322  & -- &         &5418& 85& 5.4&     &     &    &    &  5.668&0.764& * \\
131977  &5568&            &4683& 62& 6.8&     & 4605&4609&4551&  6.909&1.091& \\
135204  & -- &            &5413& 91& 4.6&     &     &    &    &  5.398&0.742& \\
135599  & -- &            &5257& 86& 5.1&     &     &    &    &  5.976&0.804& \\
137107  &5727& 2 Eta CrB  &6037& 60& 6.9&     &     &    &    &  4.237&0.507& \\
139323  & -- &            &5204& 90& 7.7&     &     &    &    &  5.909&0.943& \\
139341  & -- &            &5242& 90& 7.9&     &     &    &    &  5.115&0.898& \\
140538  &5853& 23 Psi Ser &5675&100& 3.5&     &     &    &    &  5.045&0.640& \\
141004  &5868& 27 Lam Ser &5884& 81& 4.4& 5937& 5897&    &    &  4.072&0.558& \\
143761  &5968& 15 Rho CrB &5865& 81&11.1& 5782&   &5726&    &  4.209&0.560& * \\
144287  & -- &            &5414& 93& 5.7&     &     &    &    &  5.450&0.739& \\
144579  & -- &            &5294& 89&10.3&     &     &5309&5275&  5.873&0.707& \\
145675  & -- &            &5406& 98&12.1&     &     &    &    &  5.319&0.864& \\
146233  &6060& 18 Sco     &5799& 96& 3.8&     &     &    &    &  4.770&0.614& \\
149661  &6171& 12 Oph     &5294& 90& 3.2&     &     &    &    &  5.817&0.817& \\
151541  & -- &            &5368& 88& 6.4&     &     &    &    &  5.630&0.757& \\
152391  & -- &            &5495& 82& 4.5&     &     &    &    &  5.512&0.732& \\
154345  & -- &            &5503& 87& 5.6&     &     &    &    &  5.494&0.708& \\
154931  & -- &            &5910& 82& 6.7&     &     &    &    &  3.558&0.578& \\
157214  &6458& 72 Her     &5784& 85& 9.5& 5676&     &    &    &  4.588&0.572& \\
157881  & -- &            &4035&  9& 4.5&     &     &    &4011&  8.118&1.371& \\
158614  &6516&            &5641& 98& 3.6&     &     &    &    &  4.910&0.678& \\
\hline
\end{tabular}
\end{table*}

\begin{table*}
{Table 1 (Continued)}\\
\begin{tabular}{rrrcrrccccclc}
\hline
\hline
HD/BD&HR&Name&$T_{\rm eff}$&N&$\sigma$, K&$T_{\rm eff}$&$T_{\rm eff}$&
$T_{\rm eff}$&$T_{\rm eff}$& Mv & B--V& rem \\
 &  &  & this paper  & & &EDV93&AAMR96& BLG98 & DB98 & & & \\
\hline
158633  &6518&            &5290& 83&10.7&     &     &    &    &  5.896&0.737& \\
159062  & -- &            &5414& 96& 7.9&     &     &    &    &  5.485&0.706& \\
159222  &6538&            &5834& 93& 4.0&     & 5770&5708&5852&  4.653&0.617& \\
159909  & -- &            &5749& 93& 5.6&     &     &    &    &  4.459&0.657& \\
160346  & -- &            &4983& 84& 3.9&     &     &    &    &  6.382&0.950& \\
161098  & -- &            &5617& 90& 7.3&     &     &    &    &  5.294&0.632& \\
164922  & -- &            &5392& 96& 6.0&     &     &    &    &  5.293&0.789& \\
165173  & -- &            &5505& 95& 4.7&     &     &    &    &  5.388&0.732& \\
165401  & -- &            &5877& 85& 8.5& 5758&     &    &    &  4.880&0.557& \\
165476  & -- &            &5845& 90& 5.9&     &     &    &    &  4.406&0.580& \\
166620  &6806&            &5035& 75& 5.7&     & 4947&4995&4930&  6.165&0.871& \\
168009  &6847&            &5826& 93& 4.0&     & 5781&5833&5826&  4.528&0.596& \\
170512  & -- &            &6078& 43& 9.4&     &     &    &    &  3.965&0.542& \\
171067  & -- &            &5674& 81& 6.5&     &     &    &    &  5.191&0.660& \\
173701  & -- &            &5423&104& 9.7&     &     &    &    &  5.343&0.847& \\
176841  & -- &            &5841& 92& 6.2&     &     &    &    &  4.487&0.637& \\
182488  &7368&            &5435& 82& 4.4&     &     &    &    &  5.413&0.788& \\
183341  & -- &            &5911& 85& 3.9&     &     &    &    &  4.201&0.575& \\
184385  & -- &            &5552& 87& 4.1&     &     &    &    &  5.354&0.721& \\
184768  & -- &            &5713& 94& 3.9&     &     &    &    &  4.593&0.645& \\
185144  &7462& 61 Sig Dra &5271& 79& 6.3&     & 5227&    &    &  5.871&0.765& \\
186104  & -- &            &5753& 95& 5.8&     &     &    &    &  4.621&0.631& \\
186379  & -- &            &5941& 67& 9.8&     &     &    &    &  3.586&0.512& \\
186408  &7503& 16 Cyg A   &5803& 83& 3.1&     & 5763&    &5783&  4.258&0.614& \\
186427  &7504& 16 Cyg B   &5752& 77& 3.5&  & 5767&    &5752&  4.512&0.622&* \\
187123  & -- &            &5824& 86& 5.0&  &     &    &    &  4.433&0.619&* \\
187897  & -- &            &5887& 95& 5.0&     &     &    &    &  4.521&0.585& \\
189087  & -- &            &5341& 83& 4.0&     &     &    &    &  5.873&0.782& \\
189340  &7637&            &5816& 90& 8.4&     &     &    &    &  3.920&0.532& \\
190067  & -- &            &5387&100&10.3&     &     &    &    &  5.731&0.707& \\
195005  & -- &            &6075& 51& 6.7&     &     &    &    &  4.302&0.498& \\
197076  &7914&            &5821& 75& 5.6&     & 5761&5774&5815&  4.829&0.589& \\
199960  &8041& 11 Aqr     &5878& 78& 5.9& 5813&     &    &    &  4.089&0.590& \\
201091  &8085& 61 Cyg     &4264& 17&12.4&     & 4323&    &    &  7.506&1.158& \\
201092  &8086& 61 Cyg     &3808&  5&26.4&     & 3865&    &    &  8.228&1.308& \\
202108  & -- &            &5712& 82& 7.2&     & 5635&    &    &  5.186&0.610& \\
203235  & -- &            &6071& 52& 8.4&     &     &    &    &  3.606&0.468& \\
204521  & -- &            &5809& 74&13.6&     &     &    &    &  5.245&0.545& \\
205702  & -- &            &6020& 50& 4.7&     &     &    &    &  3.839&0.513& \\
206374  & -- &            &5622& 89& 5.4&     &     &    &    &  5.304&0.674& \\
210667  & -- &            &5461& 81& 5.6&     &     &    &    &  5.470&0.800& \\
211472  & -- &            &5319& 91& 5.3&     &     &    &    &  5.835&0.802& \\
215065  & -- &            &5726& 95& 9.7&     &     &    &    &  5.131&0.594& \\
215704  & -- &            &5418& 95& 4.9&     &     &    &    &  5.500&0.795& \\
217014  &8729& 51 Peg     &5778& 92& 5.4& 5755&     &    &    &  4.529&0.615&* \\
219134  &8832&            &4900& 63& 7.9&     & 4785&    &    &  6.494&1.009& \\
219396  & -- &            &5733& 91& 5.3&     &     &    &    &  3.918&0.654& \\
220182  & -- &            &5372& 94& 4.7&     &     &    &    &  5.661&0.788& \\
221354  & -- &            &5295& 95& 5.5&     &     &    &    &  5.610&0.830& \\
+32 1561& -- &            &4950& 82& 6.2&     &     &    &    &  6.493&0.919& \\
+46 1635& -- &            &4273& 12& 4.2&     &     &    &    &  7.895&1.367& \\
Sun     & -- &            &5777&889& 0.9&     &     &    &    &  4.790&0.65 & \\
\hline
\end{tabular}
\end{table*}

\section{Results and Discussion}
Table 1 contains our final $T_{\rm eff}$ determinations for 181 MS
stars. Note that we added the 44 $K$ correction to the
initial calibrations in order to reproduce the standard 5777 $K$
temperature of the Sun. For each star we report the mean $T_{\rm
eff}$, number of the calibrations used ($N$), and the standard
error of the mean ($\sigma$). For comparison, we also provide
$T_{\rm eff}$ as determined in Edvardsson et al. (\cite{edvet93},
hereafter EDV93), AAMR96, BLG98 and DB98. Absolute magnitudes
M$_{V}$ have been computed from Hipparcos parallaxes and V
magnitudes from the Tycho2 catalogue (H{\o}g et al. \cite{tyc2})
transformed into Johnson system. $(B-V)$ are also from Tycho2.
Planet harboring stars are marked with an asterisk.

As one can see from Table 1, for the majority of stars
we get an error which is smaller than 10 $K$.
The consistency of the results derived from the ratios
of lines representing different elements is very
reassuring. It tells that our 105 calibrations
are essentially independent from micro-turbulence,
LTE departures, abundances, rotation and other individual
properties of stars.
We admit though that a small systematic error may exist for
$T_{\rm eff}$ below 5000 $K$ where we had only few standard stars.

As was already mentioned, for the first approximation we took
accurate temperatures from AAMR96, BLG98 and DB98. The comparison
of our final $T_{\rm eff}$ with those derived by AAMR96, BLG98 and
DB98 is shown in Fig. 3. As a test of the internal precision of
our $T_{\rm eff}$ we investigate the $T_{\rm eff}$ -- color
relation with the Str\" omgren index b--y, using our
determinations of $T_{\rm eff}$, and those obtained by other
authors. The results are shown in Table 2 where the rms of the
linear fit is given for each author's determination, along with
our estimate of $T_{\rm eff}$ and using common stars. In each case
the scatter of the color relation is significantly improved when
adopting our temperatures, though some residual dispersion is
still present that can be attributed to the photometric errors,
reddening and the intrinsic properties of stars (metallicity,
gravity, binarity...) to which the color indices are known to be
sensitive sensitive. The improvement is particulary spectacular in
comparison with EDV93. This proves the high quality of our
temperatures and the mediocrity of $b$--$y$ as a temperature
indicator.

\section{Conclusion}

The high-precision temperatures were derived for a set of 181
dwarfs, which may serve as temperature standards in the 4000--6150
$K$ range. These temperatures are precise to within 3--13 $K$
(median 6 $K$) for the major fraction of the sample, except for
the two outliers. We demonstrated that the line ratio technique is
capable of detecting variations in $T_{\rm eff}$ of a given star
as small as 1--5 $K$. This precision may be enough to detect star
spots and Solar-type activity cycles. Of particular interest is
the application of this method to testing ambiguous cases of
low-mass planet detection,  since planets do not cause temperature
variations, unlike spots.

The next step will be the adaptation of this method to a wider range of
spectral types and for an automatic pipeline analysis of large spectral
databases.

\begin{acknowledgements}
V.K. wants to thank the staff of Observatoire de Bordeaux for the
kind hospitality during his stay there. The authors are also
grateful to the anonymous referee for the careful reading of the
manuscript and the numerous important remarks that helped to
improve the paper.
\end{acknowledgements}

\end{document}